\documentclass[aps,apl,preprint,groupedaddress,showpacs]{revtex4}
\usepackage{graphicx, color}

\begin{document}
\title{DNA-psoralen: single-molecule experiments and first principles calculations }
\author{M. S. Rocha$^1$}
\author{A. D. L\'ucio$^2$}
\author{S. S. Alexandre$^3$}
\author{R. W. Nunes$^3$}
\author{O. N. Mesquita$^3$}

\affiliation{{$^1$Departamento de F\'\i sica, Universidade Federal
de Vi\c{c}osa, \\ CEP 36570-000, Vi\c{c}osa, MG, Brazil}}

\affiliation{{$^2$Departamento de Ci\^encias Exatas, Universidade
Federal de Lavras, \\ Caixa Postal 3037, CEP 37200-000, Lavras, MG,
Brazil}} \date{\today}

\affiliation{{$^3$Departamento de F\'\i sica, ICEx, Universidade
Federal de Minas Gerais, \\ Caixa Postal 702, CEP 31270-901, Belo
Horizonte, MG, Brazil}}

\begin{abstract}
The authors measure the persistence and contour lengths of
DNA-psoralen complexes, as a function of psoralen concentration, for
intercalated and crosslinked complexes. In both cases, the
persistence length monotonically increases until a certain critical
concentration is reached, above which it abruptly decreases and
remains approximately constant. The contour length of the complexes
exhibits no such discontinuous behavior. By fitting the relative
increase of the contour length to the neighbor exclusion model, we
obtain the exclusion number and the intrinsic intercalating constant
of the interaction. {\it Ab initio} calculations are employed in
order to provide an atomistic picture of these experimental
findings.
\end{abstract}

\pacs{87.80.Cc, 87.14.gk, 87.15.La}

\maketitle

DNA interactions with ligands such as drugs or proteins have been
extensively studied and reviewed by many authors in the past years
\cite{Fritzsche, Tessmer, Sischka, Sinden, Rocha, RochaJCP2, Cimino,
McCauley2}. Some of these drugs, like daunomycin and ethidium
bromide (EtBr), exhibit intercalative binding \cite{RochaJCP2,
Fritzsche,Sischka}. Psoralen, by contrast, presents other forms of
linkage to the DNA bases. In fact, it is well known that this drug
can absorb photons and form covalent bonds with the DNA bases, when
the sample is illuminated with ultraviolet-A (UVA) light ($\lambda$
= 320-400 nm) \cite{Sinden}. Psoralens are compounds from the family
of furocoumarins, broadly used in the medical treatment of various
skin diseases like psoriasis, vitiligo, and some other kinds of
dermatitis \cite{McNeely}.

\begin{figure}[h]
\includegraphics[width=0.7\columnwidth]{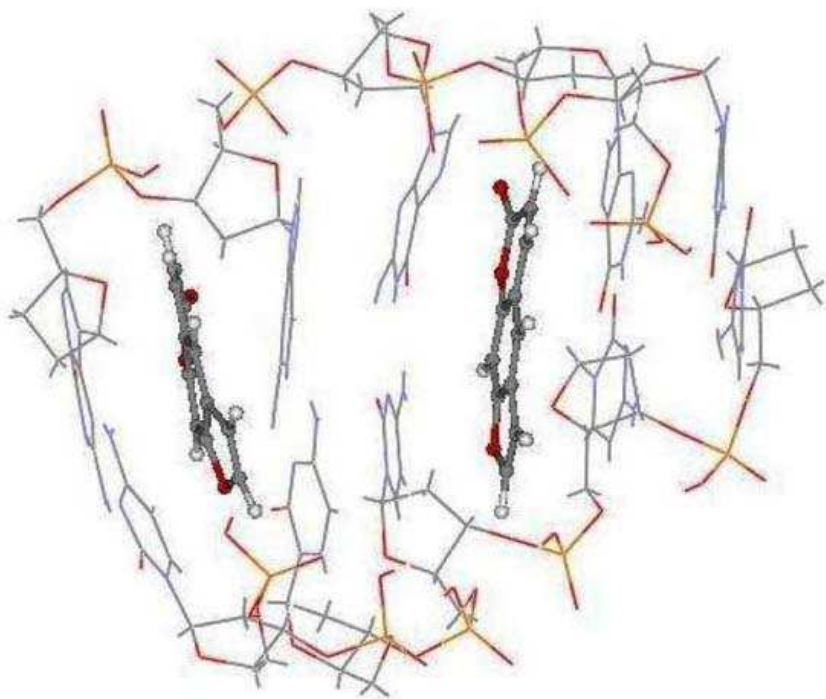}
\caption{Psol-DNA fragment with five base CG pairs and two
intercalated psoralens obtained from our {\it ab initio} DFT
calculations.} \label{estrut}
\end{figure}
Although psoralens have been extensively used in medicine for many
years, the interaction of the drug with DNA is not yet completely
understood. In this work, we study the intercalative binding and
crosslink formation in DNA-psoralen complexes (Psol-DNA), in order
to clarify the quantitative aspects of the interaction between
psoralen and DNA. We are interested in probing the modifications of
the mechanical properties of DNA molecules when interacting with
psoralen. The elasticity and size changes of these molecules, as a
function of the total psoralen concentration ($C_P$), are measured
by performing single-molecule stretching experiments, which give
information about persistence ($\xi$) and contour ($L$) lengths of
Psol-DNA. The total psoralen concentration is the sum of the
intercalated concentration ($C_i$) and the free concentration in
solution of psoralen. We also define the intercalated psoralen
fraction $r$=$C_i$/$C_{bp}$, where $C_{bp}$ is the DNA base-pair
concentration, fixed in all experiments at 11 $\mu$M. In order to
elucidate the microscopic mechanisms behind the psolaren-DNA
interaction, we also perform first principles calculations for
Psol-DNA fragments (an example is shown in Fig.~\ref{estrut}) to
determine the behavior of the Psol-DNA stiffness (which is related
to the Young modulus) as a function of $C_P$, and also to obtain the
maximum fraction of intercalated psoralen into DNA, i.e., the
limiting value of $C_P$ at which a Psol-DNA complex is still
structurally stable.

Experimentally, we observe that as $C_P$ increases the Psol-DNA
persistence length initially increases and then undergoes an abrupt
transition to a lower value at a critical total psoralen
concentration ($C_P^{crit}$), a behavior we also observe in UV-light
illuminated complexes. In the latter case, the Psol-DNA complex
attains higher persistence length values for $C_P$ $<$ $C_P^{crit}$,
which is indicative of crosslinking, where the psoralen binds
covalently preferably to thymine bases in opposite DNA strands, upon
illumination. As discussed below, the {\it ab initio} calculations
are in qualitative agreement with these experimental findings,
enabling us to analyze the changes in the molecular structure that
are responsible for the experimentally observed behavior.

\begin{figure}[h]
\includegraphics[width=0.7\columnwidth]{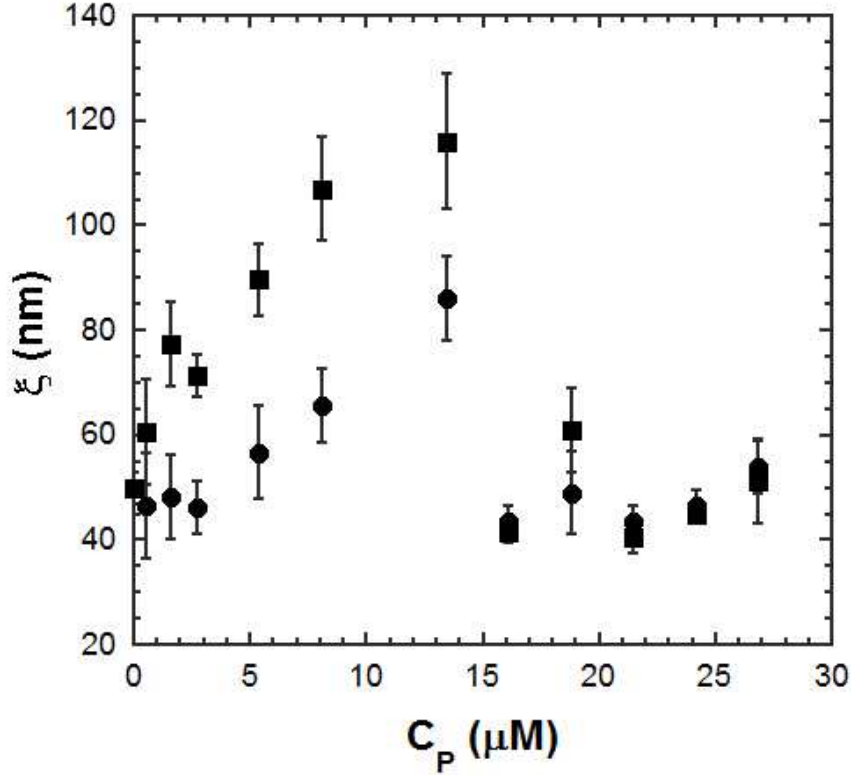}
\caption{Persistence length $\xi$ (in nm) as a function of total
psoralen concentration $C_P$ (in $\mu$M) for Psol-DNA complexes.
{\textit Circles}: intercalated Psol-DNA; {\textit Squares}:
crosslinked Psol-DNA. The DNA base-pair concentration used was
$C_{bp}$=11 $\mu$M.} \label{AxC}
\end{figure}
The intercalated complexes are obtained by simply waiting for the
psoralen to intercalate the $\lambda$-DNA in the sample. The
crosslinked complexes are obtained by illuminating the sample with a
UVA mercury lamp. We use an optical tweezers to trap a polystyrene
bead attached to the DNA molecule, while pulling the microscope
stage with a controlled velocity. We then make force $\times$
extension curves, and use the approximate expression derived by
Marko and Siggia~\cite{Marko} to obtain the persistence and contour
lengths of the bare DNA and Psol-DNA complexes. The persistence
length $\xi = YI/kT$ ($Y$ is the Young modulus, $I$ is the
geometrical moment of inertia of the polymer, and $kT$ is the
thermal energy) is a mesoscopic quantity which depends on the local
elasticity of DNA. Its value is $\xi\sim$ 50 nm for bare DNA under
physiological conditions. In our experiments, the forces applied to
stretch the DNA are within the entropic force regime (forces
$\lesssim$ 3 pN), thus avoiding externally distorting the structure
and shifting the DNA/psoralen chemical equilibrium (enthalpic
effects). The details about our experimental setup, experimental
procedure, and optical-tweezers calibration can be found in
Refs.~\cite{Rocha, Nathan1, RochaJCP2}.

In an attempt to understand the behavior of $\xi$ as a function of
$C_P$, we perform first principles calculations of the structural
stability and stiffness changes produced by intercalation of
psoralen into fragments of dry poly(dG)-poly(dC) in its acid form.
We investigate DNA models with fragments consisting of four and five
guanine-citosine (CG) pairs. A five-base-pair fragment with two
intercalated psoralens is shown in Fig.~\ref{estrut}. Our first
principles density functional theory (DFT) simulations use the
SIESTA code~\cite{soler}. Exchange and correlation effects are
described within the generalized gradient
approximation~\cite{perdew}. A double-$\zeta$ basis set of atomic
orbitals of finite range is used, with polarization functions added
for phosphorus and for the atoms involved in the hydrogen bridges.
This methodology has been shown to provide a good description of the
DNA structure~\cite{pablo,parrinelo,mdna}. We treat isolated finite
fragments in a periodic supercell with large vacuum regions. For the
DNA bases at the edges of the fragments, one atom was held fixed in
the same position as in the infinite periodic DNA~\cite{pablo}.
Broken bonds at the edges are saturated with hydrogen atoms. Full
geometry relaxations were performed (forces $< 0.04~$eV/\AA). The
stability of Psol-DNA complexes is addressed by computing their
formation energies defined as the difference between the total
energy of the Psol-DNA complex and the sum of the total energies of
the bare-DNA fragment and the isolated psoralen molecule, after full
relaxation of the initial geometries.

\begin{figure}[h]
\includegraphics[width=0.7\columnwidth]{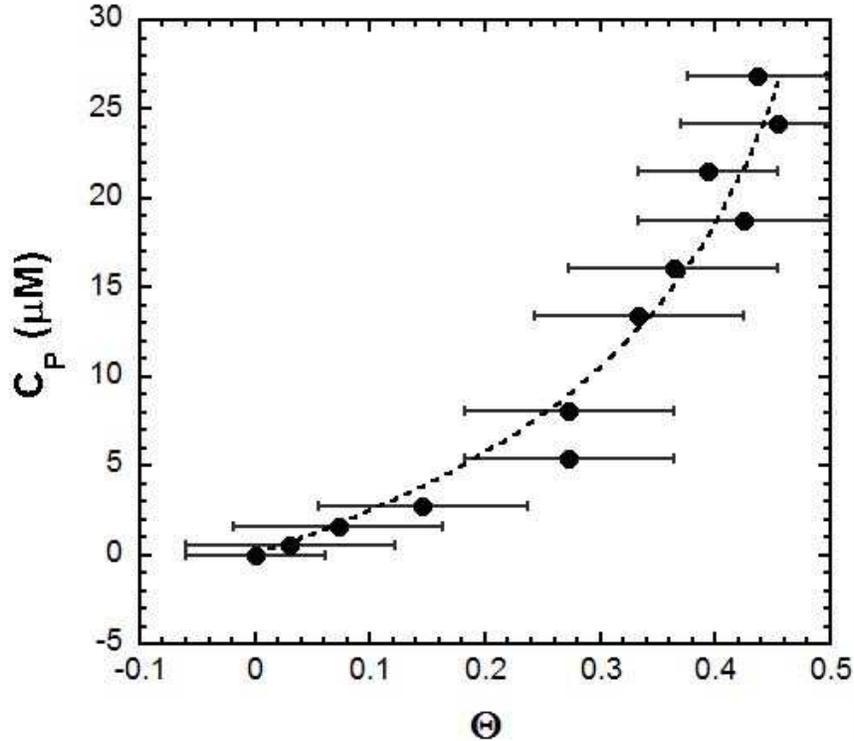}
\caption{Total psoralen concentration $C_P$ as a function of
relative increase of the contour length $\Theta$ for Psol-DNA
intercalated complexes. $\Theta$ for psoralen molecules equals the
intercalated psoralen fraction $r$. Dots are the experimental data
and the dashed line is a fit using Eq. 1.} \label{bindingPSO}
\end{figure}
In Fig.~\ref{AxC}, we plot the measured values of $\xi$ as a
function of $C_P$, for both intercalated and crosslinked Psol-DNA
complexes. The figure shows that $\xi$ increases monotonically until
the critical concentration $C_P^{crit}\sim 13.4~\mu$M is reached.
For $C_P> C_P^{crit}$, $\xi$ decays abruptly to around 50 nm, and
remains approximately constant at this value, at least for the
concentration range in our experiments. Note that $C_P^{crit}$ is
the same for both intercalated and crosslinked complexes. A similar
abrupt transition of the persistence length was recently reported by
two of us for other intercalating drugs (daunomycin and ethidium
bromide)~\cite{RochaJCP2}. The fact that Psol-DNA complexes exhibit
this same behavior is a strong evidence that this transition is of
general character for intercalating molecules, at least in the
low-force regime ($\lesssim$ 3 pN) used in our experiments. This
abrupt transition may be caused by local formation of denaturing
bubbles, as proposed in Refs.~\cite{Harris} and~\cite{RochaPB},
which softens the structure of the Psol-DNA complex. In
Ref.~\cite{Harris} large scale molecular dynamics calculations
indicate the formation of denaturing bubbles when pure DNA is
stretched above 65 pN forces. Since we use very small stretching
forces it is likely that the abrupt transition we observe is caused
by some intrinsic DNA structural changes induced by the
intercalating drug.

In Fig.~\ref{bindingPSO} we show the relative increase of the
contour length, $\Theta=(L-L_0)/L_0$, as a function of $C_P$ for the
same intercalated complexes shown in Fig.~\ref{AxC}, where $L_0$ is
the contour length for $C_P=0$. Note that the contour length does
not exhibit any abrupt changes. The dashed line in this figure is a
fit using Eq.1 below, derived from the neighbor exclusion model
(NEM)~\cite{McGhee} (see Ref. \cite{RochaJCP2} for details).
Considering that each intercalated psoralen molecule increases the
length of DNA by the base-pair distance 0.34~nm~\cite{Sinden},
$\Theta$ is equal to the intercalated psoralen fraction $r$. In this
case the NEM gives
\begin{equation}
C_P = {C_{bp}}\Theta + \frac{\Theta(1 - n\Theta +
\Theta)^{n-1}}{K_i(1 - n\Theta)^n} \label{nem},
\end{equation}
where $n$ is the exclusion number and $K_i$ is the intrinsic
intercalating constant. From the fitting we determine the parameters
$n = 1.43 \pm 0.13$ and $K_i = (8.8 \pm 2.4)\times 10^4$ M$^{-1}$,
for intercalative binding of psoralen. The large error bars for
$\Theta$ are due to the fact that different DNAs were used, and
there is a natural length distribution for $\lambda$-DNA. The abrupt
transition for $\xi$ occurs around $\Theta=$0.38, while the maximum
psoralen intercalated fraction ($1/n$) is between 0.64 and 0.77.

\begin{figure}[h]
\includegraphics[width=0.8\columnwidth]{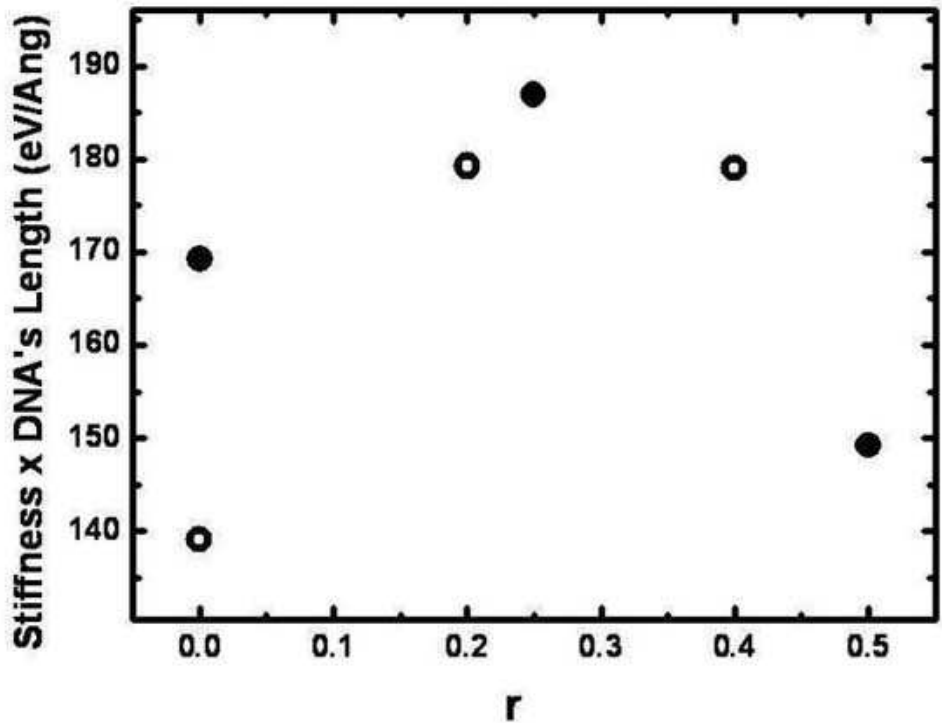}
\caption{Stiffness as a function of intercalated psoralen fraction
($r$), from {\it ab initio} DFT calculations. Results for $r =
0,~0.25,~{\rm and}~ 0.50$ were obtained using a four-base-pair
model. Results for $r = 0,~ 0.20,~ 0.40,~{\rm and} ~0.60$ were
obtained using a five-base-pair model.} \label{teoria}
\end{figure}
Figure~\ref{teoria} shows the behavior of the stiffness, which is
proportional to the Young modulus (for fixed DNA length), as a
function of $r$, computed using the DFT methodology. Note that the
stiffness increases as $r$ increases, and an abrupt decrease is
observed for $r$ between 0.40 and 0.50, in reasonably good agreement
with the data for the persistence length of Fig~\ref{AxC}, where
this transition occurs around $\Theta=r=$0.38. Our calculations
indicate that before the transition two psoralens are well
intercalated into the DNA structure, while after the transition one
of the psoralens binds with the citosine, and the hydrogen bond
between the two bases is broken, indicating that local denaturation
may indeed be the mechanism behind the abrupt change in stiffness
and consequently in the persistence length. Moreover, our
calculations also indicate that when $r$ increases above 0.60, the
intercalated structure is unstable, which sets an upper bound on the
maximum fraction of intercalated psoralen within this theoretical
model, in good agreement with the lower-limit value of 0.64 measured
experimentally. Our interpretation for the microscopic nature of the
transition rests on the good qualitative and quantitative agreement
between theory and experiment. We must, however, not refrain from
stating the limitations of our model system, since calculations in
such small fragments are strongly affected by boundary effects. This
is reflected in the large values we obtain for the theoretical
stiffness, which are also due to the fact that such small fragments
are much stiffer than the much longer DNA chains in the experiments
(The stiffness $\times$ length we compute for bare DNA using the
four-base-pair model is about 20\% higher than that of the
five-base-pair model.)

In conclusion, the authors observe experimentally that the
persistence length of DNA-psoralen complexes increases with psoralen
concentration up to a critical concentration, decreasing abruptly
and remaining approximately constant above this concentration, a
behavior also observed in UV-light illuminated crosslinked
complexes. {\it Ab initio} calculations show a similar behavior for
the stiffness of psoralen-DNA complexes, and also indicate that
local denaturation and binding between psoralen and a DNA base may
be the mechanism behind this persistence length transition.

The authors acknowledge support from the European Grant EU
(FP6-029192), and from the Brazilian agencies: FAPEMIG, CAPES, CNPq,
Instituto do Mil\^enio de Nanotecnologia/MCT and Instituto do
Mil\^enio de \'Optica N\~ao-linear, Fot\^onica e Biofot\^onica/MCT. \\

\end{document}